\documentclass[11pt,twoside]{article}  
\usepackage{adassconf}

\begin{document}   

%
%

\paperID{P1.1.8}

%
%
%
%

\title{The Multimission Archive at the Space Telescope 
Science Institute in the context of VO activities}
\titlemark{MAST in the context of VO activities}

%

\author{Inga Kamp\altaffilmark{1}, Randall Thompson\altaffilmark{1},
        Alberto Conti\altaffilmark{1}, Dorothy Fraquelli\altaffilmark{1}, 
        Tim Kimball\altaffilmark{1}, Karen Levay\altaffilmark{1}, 
        Bernie Shiao\altaffilmark{1}, Myron Smith\altaffilmark{1}, 
        Rachel Somerville\altaffilmark{1}, 
        and Richard L. White\altaffilmark{1}}
\affil{Space Telescope Science Institute, 3700 San Martin Drive, Baltimore,
       MD 21218}



\contact{Inga Kamp}
\email{kamp@stsci.edu}

%
%

\paindex{Kamp, I.}
\aindex{Thompson, R.}
\aindex{Conti, A.}
\aindex{Fraquelli, D.}
\aindex{Kimball, T.}
\aindex{Levay, K.}
\aindex{Shiao, B.}
\aindex{Smith, M.}
\aindex{Somerville, R.}
\aindex{White, R.L.}

%
%

\authormark{Kamp et al.}




\begin{abstract}          
In the past year, the Multimission Archive at the Space Telescope 
Science Institute (MAST) has taken major steps in making MAST's holdings 
available using VO-defined protocols and standards, and in implementing 
VO-based tools. For example, MAST has implemented the Simple Cone 
Search protocol, and all MAST mission searches may be returned in the 
VOTable format, allowing other archives to use MAST data for their VO 
applications. We have made many of our popular High Level Science 
Products available through Simple Image Access Protocol (SIAP), and are 
implementing the VO Simple Spectral Access Protocol (SSAP). The cross 
correlation of VizieR catalogs with MAST missions is now possible, and 
illustrates the integration of VO services into MAST. The user can 
easily display the results from searches within MAST using the 
plotting tool VOPlot. MAST also participates in the NVO registry service. 
Thus, the user can harvest MAST holdings simultaneously with data from 
many other surveys and missions through the VO DataScope Data Inventory 
Service.

\end{abstract}


\section{Introduction}

The Multimission Archive at Space Telescope Science Institute is NASA's 
UV-optical science archive center. It contains the Hubble Space Telescope archive 
plus more than a dozen additional active, planned, and legacy mission datasets. 
The total MAST data volume exceeds 20 Terabytes, making it one of the most 
significant astronomical collections available on-line.  

MAST is an important node in the upcoming virtual observatory and we are 
continually improving our data discovery and retrieval abilities. MAST has 
taken major steps in making its holdings available using VO-defined protocols 
and standards, and in implementing VO-based tools. All MAST mission searches 
may be returned in the VOTable format, allowing other archives to use MAST 
data for their VO applications.

\section{MAST Webservices}

\begin{figure}
\epsscale{0.72}
\plotone{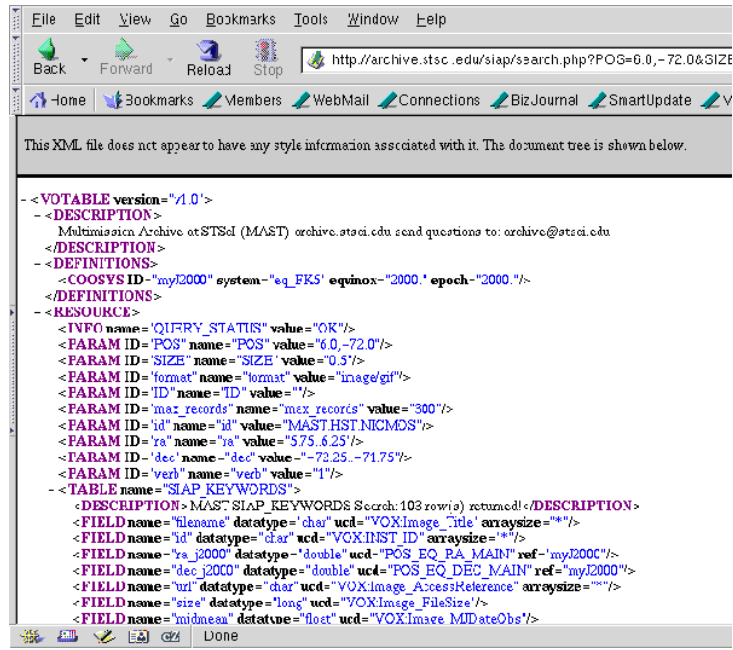}
\caption{Example of the Simple Image Access Protocol: the command 
{\bf \sl http://archive.stsci.edu/siap/search.php?POS=6.0,-72.0\&SIZE=0.5\&format=gif\&ID=nicmos}
returns the above VOTable containing the information on all matching datasets.} \label{SIAP}
\end{figure}

Currently, MAST provides three different  forms of webservices, HTTP GET 
Requests, an RSS news service and dataset verifier SOAP service. 
Additional information can be found at 
{\bf \sl http://archive.stsci.edu/vo/mast\_services.html}.
HTTP GET Requests allow search parameters to be included in the URL. 
As such, they can be called from within programs to automate data searches. 
Currently, MAST provides this service for Mission Searches, Simple Cone 
Searches and the Simple Image Access Protocol. The results are returned 
in VOTable XML format. The result of the Mission Searches can also be
returned as excel spreadsheet, and comma-separated values which can 
simplify ingesting results into user-written programs. 
Fig.~\ref{SIAP} shows the VOTable returned
for a positional search in the archived NICMOS images: the search radius
is 0.5 arcsec and the search is restricted to data in gif-format.

The RSS news service offers to search a list of MASTs most recent news items and returns
the result in RSS XML-format. The output can also be displayed in HTML.

\begin{figure}[ht]
\vspace*{3mm}
\epsscale{1.0}
\plotone{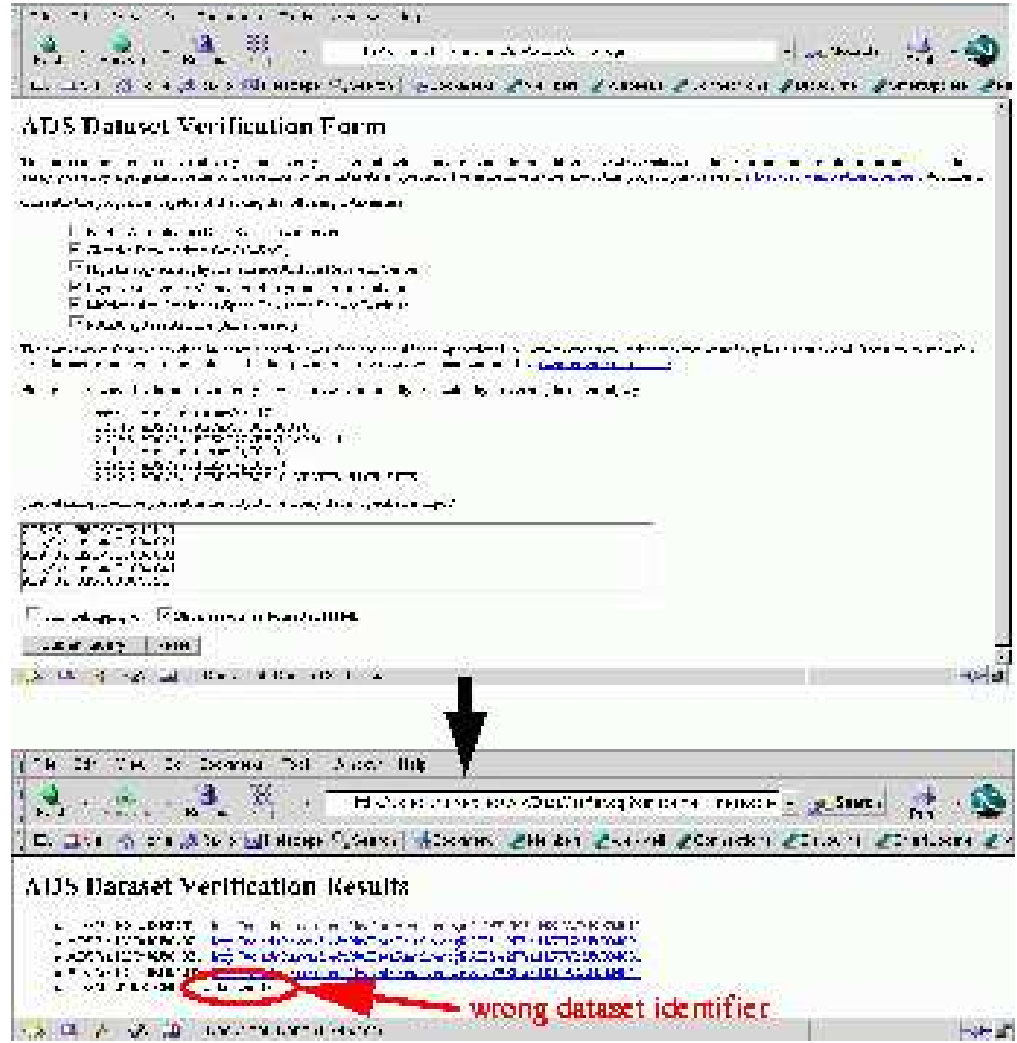}
\caption{Example of the SOAP-based web service: enter the dataset names in
the verification form and submit the request. The result contains the verification
of the dataset name as well as links to these datasets.} \label{SOAP}
\end{figure}

MAST has one SOAP-based web service, which is accessed via the ADS. The web service allows 
users to enter a data set name and obtain a verification and link to the archival website
of this dataset. More information about the ADEC naming conventions can be found at
{\bf \sl http://archive.stsci.edu/pub\_dsn.html}. The service can be accessed from the 
web form at:\\
{\bf \sl http://ads.harvard.edu/ws/DataVerifier},
or, using the latest ADEC naming conventions:\\
{\bf \sl http://vo.ads.harvard.edu/dv/DataVerifier.cgi} (Fig.~\ref{SOAP})\,\,.

\section{VOPlot at MAST}

\begin{figure}
\epsscale{.60}
\plotone{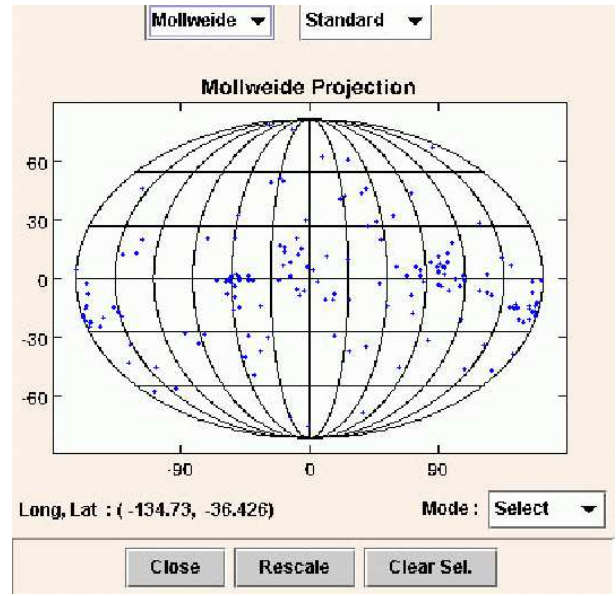}
\caption{The galactic distribution of STIS high-resolution UV observations: the
graphical display tool VOPlot is called from the result page of a 
MAST archive query.} \label{VOPlot}
\end{figure}

Users may plot search results from any MAST mission or VizieR catalog search with 
the new JAVA-based graphical display tool called VOPlot (developed within the Indian 
VO project in collaboration with CDS). The example in Fig.~\ref{VOPlot} shows the 
galactic distribution of STIS UV observations with the Echelle gratings.

\section{MAST and VizieR}

\begin{figure}
\epsscale{0.9}
\plotone{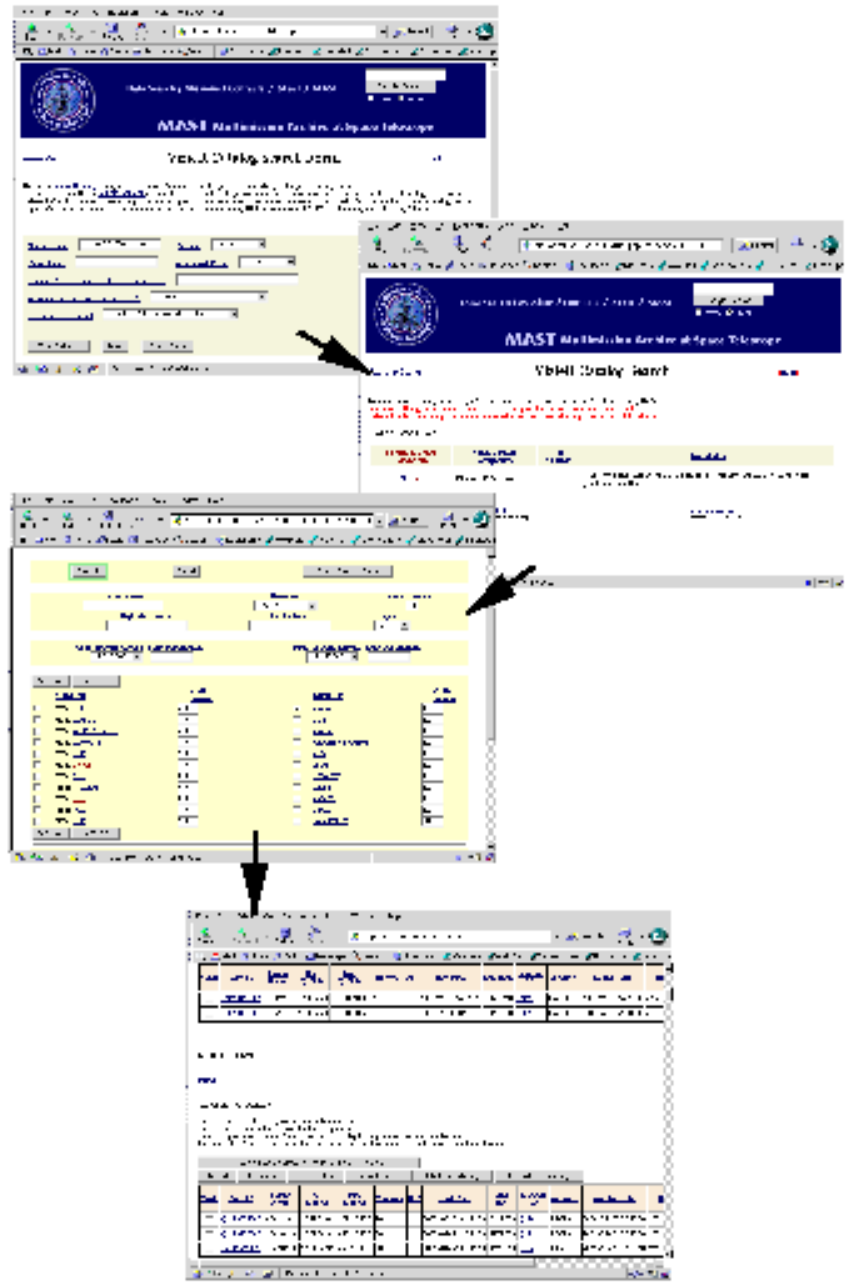}
\caption{Provide the catalog information for the cross correlation at
{\bf \sl http://archive.stsci.edu/vizier.php}. After choosing the cross 
correlation option for the catalog of Vega-type stars, the FUSE mission 
was selected and the last page shows a pane of the resulting web table.} \label{VizieR}
\end{figure}

Users may now search the entire set of 4\,000 VizieR catalogs and cross correlate 
the results with any MAST mission. This uses a VizieR web service which communicates 
using the VOTable standard. The example shows the result of a cross correlation between 
the Vega-type star catalogue and archived FUSE observations. Enter the catalog information 
at {\bf \sl http://archive.stsci.edu/vizier.php} and cross correlate
it with the FUSE data in MAST. Fig.~\ref{VizieR} shows how to navigate through
the cross correlation. In addition, it provides a cutout of the returned 
information on the available FUSE datasets for Vega-type stars from the 
specified catalog.

\end{document}